\documentclass[12pt]{article}

\usepackage{amsmath}
\usepackage{amsfonts}
\usepackage[nosort]{cite}

\numberwithin{equation}{section}

\def\beq#1\eeq{\begin{equation}#1\end{equation}}
\def\bea#1\eea{\begin{align}#1\end{align}}

\newcommand{\del}{\partial}
\newcommand{\class}{\mathcal{C}}

\newcommand{\R}{\mathbb{R}}
\newcommand{\Z}{\mathbb{Z}}
\newcommand{\C}{\mathbb{C}}

\newcommand{\bra}[1]{\langle #1|}
\newcommand{\ket}[1]{|#1\rangle}
\newcommand{\rbra}[1]{(#1|}
\newcommand{\rket}[1]{|#1)}
\newcommand{\dbra}[1]{\langle\!\langle #1|}
\newcommand{\dket}[1]{|#1 \rangle\!\rangle}

\newcommand{\cor}[1]{\langle #1 \rangle}

\newcommand{\trace}[1]{\,\, \underset{%
  \mbox{$\scriptscriptstyle #1$}}{\mathrm{Tr}}\,\,}

\newcommand{\alphat}{\tilde{\alpha}}

\newcommand{\Jt}{\tilde{J}}
\newcommand{\Lt}{\tilde{L}}

\newcommand{\delbar}{\bar{\partial}}
\newcommand{\zbar}{\bar{z}}
\newcommand{\wbar}{\bar{w}}
\newcommand{\gammabar}{\bar{\gamma}}
\newcommand{\betabar}{\bar{\beta}}

\newcommand{\Ishibashi}[1]{\dket{#1}_{\!I}}
\newcommand{\Cardy}[1]{\dket{#1}_{\!C}}

\begin{document}
\baselineskip=6mm

\begin{titlepage}
\nopagebreak
\vskip 5mm
\begin{flushright}
hep-th/0007141 \\
TU-595
\end{flushright}

\vskip 30mm
\begin{center}
{\Large Free Field Realization of $D$-brane in Group Manifold}
\vskip 15mm
Hiroshi \textsc{Ishikawa} and
Satoshi \textsc{Watamura}
\vskip 5mm
\textsl{%
Department of Physics, Tohoku University \\
Sendai 980-8578, JAPAN\\
}
\texttt{%
\footnotesize
ishikawa, watamura@tuhep.phys.tohoku.ac.jp
}
\end{center}
\vskip 15mm

\begin{quote}
We construct the boundary state for the $D$-brane
in the $SU(2)$ group manifold directly in terms of the group
variables.
We propose a matching condition for the left- and the 
right-moving sectors including the zero modes 
that describes a $D$-brane of the Neumann-type.
The free field realization of the WZW model 
is used to obtain the boundary state subject to 
the matching condition.
We show that the resulting state coincides with Cardy's state.
The structure of the BRST cohomology is realized by imposing
the invariance of the state under the Weyl group
of the current algebra.
\end{quote}

\vfill
\end{titlepage}

\section{Introduction}
Various properties of
$D$-branes in flat backgrounds are now well understood
from the sigma model approach as well as from
the exact conformal field theory (CFT) approach. 
In the former one, 
a $D$-brane is a submanifold of the target space
on which the boundary of the worldsheet is lying.
By construction,
geometrical data such as the position and the shape
of the brane are obvious.
In the latter case, we have a boundary CFT and
a $D$-brane is described by the boundary state.
The geometrical data of branes are encoded in the boundary states
and can be retrieved by acting the string coordinate on the states.

In the case of curved backgrounds, however, this correspondence
between the geometry of $D$-branes and the boundary states
is not so straightforward since we do not know the precise
relation between the exact CFT and the target space.
Even if we can construct a boundary state subject to an appropriate
boundary condition, we are left with the problem to
extract the geometrical information from the resulting state. 

The $D$-branes in group manifolds have recently attracted 
much attention \cite{KS,KO,AS,S,ARS,FFFS,BDS,ARS2,T,S2,AS2}.
They give us an example of curved branes in the non-trivial 
backgrounds.
The corresponding worldsheet theory is the Wess-Zumino-Witten (WZW)
model and is exactly solvable. We can therefore take both approaches
of the sigma model and the exact CFT to analyze the branes.
A group manifold is a good example to study
the geometrical interpretation of $D$-branes.
\footnote{%
There are also several works studying the case of
the Calabi-Yau manifold \cite{OOY,RS,GS,BDLR}.
}

In the WZW model on the group manifold $G$, 
we have the $G \times G$ symmetry of the left-
and the right-translations.
Correspondingly, we have the left- and the right-moving current,
$J(z)$ and $\Jt(\zbar)$, respectively.
The gluing condition that preserves the diagonal part of
$G \times G$ takes the following form
\beq
  J + \Jt = 0 \, .
\eeq
This condition implies that the worldvolume of 
the $D$-brane coincides with a conjugacy class of $G$ \cite{AS,G}.
In exact CFT, one can solve the same problem to obtain the
boundary states satisfying the above condition
\cite{Ishibashi,Cardy}. 
Since these two approaches treat the same object, 
we expect that it is possible to relate the boundary states 
with the conjugacy class of $G$.
Although there is an argument \cite{G,FFFS} 
based on the wave function of the string zero modes,
a more direct correspondence including the oscillator modes
is desirable.

In this paper, we study the above problem of the relation between
the geometry of $D$-branes and the boundary states 
for the case of $G = SU(2)$, focusing on
the construction of boundary states in terms of group variables.
We treat directly the left- and the right-moving group 
variables, $g_L(z)$ and $g_R(\zbar)$, rather than the currents
$J(z)$ and $\Jt(\zbar)$.
By using the group variables, the geometrical
meaning of the boundary state is manifest. 

This paper is organized as follows.
In the next section, we review some of the known results about
the $D$-brane in the $SU(2)$ group manifold. 
We also discuss the structure of the wave function 
for the boundary states found by Cardy \cite{Cardy}.
In Section 3, we propose the matching condition of the left- 
and the right-moving group variables corresponding to
the $D$-brane wrapped around the conjugacy class.
We then construct the boundary states which satisfy the matching 
condition using the free field realization \cite{GMOM,G2,BF}.
We require invariance of the states under the Weyl group
of the current algebra, and show that the resulting states
coincide with Cardy's states.
This fact clarifies the geometrical meaning of Cardy's states.
The final section is devoted to discussions.

\section{$D$-brane in the $SU(2)$ group manifold}
\label{review}
In this section, we briefly review some properties
of the $D$-branes in the WZW model.

We first recall some basic facts about the functions
on a group manifold.
Let $(\pi, V_\pi)$ be a unitary irreducible representation of 
a compact group $G$
and take $\{\ket{m}\}$ to be an orthonormal basis of $V_\pi$.
Then, the matrix element $\pi_{mn}$ is written as
\bea
  \pi_{mn}(g) &= \bra{m} \pi(g) \ket{n} \, , \quad g \in G \, ,\\
\intertext{or equivalently,}
  \pi(g) \ket{n} &= \ket{m} \pi_{mn}(g) \, .
  \label{basis}
\eea
The set $\{\pi_{mn}\}$ form a basis of 
$L^2(G)$, the space of all square integrable functions on $G$. 
They are orthogonal to each other 
\footnote{Here, $\overline{\pi}$ stands for 
the complex conjugate of $\pi$.},
\beq
  \int_G \!dg\, \overline{\pi_{ij}(g)} \pi'_{kl}(g) = 
  \begin{cases}
     0 & 
	 \text{if $\pi \ne \pi'$}, \\
	 \frac{1}{\text{dim}\, \pi} \delta_{ik} \delta_{jl} &
	 \text{if $\pi = \pi'$},
  \end{cases}
  \label{Schur}
\eeq
which is known as Schur's orthogonality relation.

We denote the conjugacy class including an element $t$ as $\class(t)$,
\beq
  \class(t) = \{g t g^{-1}\, , \, g \in G\} \, .
  \label{class}
\eeq
A function $\psi(x)$ invariant under the adjoint action 
$x \rightarrow g x g^{-1}$ of $G$ is called a class function,
since it is constant along the conjugacy class.
The character $\chi_\pi(g)$ of the representation $\pi$ is defined as
\beq
  \chi_\pi(g) = \trace{V_\pi} \pi(g) = \sum_m \pi_{mm}(g) \, .
\eeq
One can take $\chi_\pi$ as an orthonormal basis of the class functions.
In fact, 
\beq
  \int_G \!dg\, \overline{\chi_\pi(g)} \chi_\pi'(g) = 
    \delta_{\pi \pi'} \, ,
\eeq
which follows from \eqref{Schur}.

The $\delta$-function $\delta(g, g')$ on $G$ is characterized by
the equation
\beq
  f(g) = \int_G \! dg' \, \delta(g, g') f(g') \, , \quad
  f \in L^2(G) \, .
\eeq
We can write down the $\delta$-function in terms of $\pi_{mn}$,
\beq
  \delta(g, g') = 
  \sum_{\pi; m, n} (\text{dim}\, \pi) \,
  \pi_{mn}(g) \overline{\pi_{mn}(g')} =
  \sum_{\pi} (\text{dim}\, \pi) \, \chi_\pi(g {g'}^{-1}) \, , 
\eeq
where the sum is over all the unitary irreducible representations 
of $G$.
Making a superposition of $\delta$-functions
over the conjugacy class $\class(t)$,
we can also write the $\delta$-function $\delta_t$ that 
concentrates on $\class(t)$,
\beq
  \delta_t(g) = \int_G \! dg' \, \delta(g' t {g'}^{-1}, g)
  = \sum_\pi \chi_\pi(t) \overline{\chi_\pi(g)} \, .
  \label{delta}
\eeq

From the left- and the right-translation on $G$,
we obtain a natural action of $G \times G$ on $L^2(G)$,
\beq
\begin{aligned}
  (L_g \psi)(x) &= \psi(g^{-1} x) \, , \\
  (R_g \psi)(x) &= \psi(x g) \, .
  \label{LR}
\end{aligned}
\eeq
Taking $\psi = \overline{\pi}_{mn}$, we obtain
\beq
\begin{aligned}
  (L_g \overline{\pi}_{mn})(x) &= 
    \overline{\pi}_{ln}(x) \pi_{lm}(g) \, , \\
  (R_g \overline{\pi}_{mn})(x) &= 
    \overline{\pi}_{ml}(x) \overline{\pi}_{ln}(g) \, ,
	\label{GxG}
\end{aligned}
\eeq 
which means that
$\overline{\pi}_{mn}$ transforms as $\pi \times \overline{\pi}$ of
$G \times G$. 
$L^2(G)$ is therefore decomposed in the following form
\beq
  L^2(G) \cong \bigoplus_i 
  V_{\pi_i} \otimes V_{\overline{\pi_i}}\, ,
  \label{L2G}
\eeq
where the sum is over all the unitary irreducible representations 
of $G$.

The wave function of a particle on $G$ is an element
of $L^2(G)$. 
Hence, the Hilbert space of the quantum mechanics on $G$
is identified with $L^2(G)$ and has the structure of \eqref{L2G}.
Using the basis given in \eqref{basis}, we can write a basis of
the Hilbert space in the form of $\ket{m} \otimes \overline{\ket{n}}$.
From the transformation property \eqref{GxG} of $\pi_{mn}$,
we can identify $\ket{m} \otimes \overline{\ket{n}}$ with
the wave function $\overline{\pi}_{mn}$, 
\beq
  \overline{\pi}_{mn} \longmapsto 
   \frac{1}{\sqrt{\text{dim }\pi}} 
   \ket{m} \otimes \overline{\ket{n}}\, ,
   \quad \ket{m}, \ket{n} \in V_\pi \, .
   \label{map}
\eeq
The factor $1/\sqrt{\text{dim }\pi}$ is necessary to
preserve the inner product of \eqref{Schur}. 

\vspace{.5\baselineskip}
Let us consider the case of string theory.
In the WZW model, we have two chiral currents 
$J$ and $\Jt$ corresponding to the left- and the right-translation 
on $G$:
\beq
\begin{aligned}
  J(z)   &= \sum_n J^a_n T^a \, z^{-n-1} 
          = - k\, \del g \, g^{-1} \, ,   \\
  \Jt(\zbar) &= \sum_n \Jt^a_n T^a \, \zbar^{-n-1}
              = k\, g^{-1} \delbar g \, .  
\end{aligned}
\label{J}
\eeq
Here we parametrize the worldsheet by the coordinates 
$(z, \zbar)$, and $k$ is the level of the model.
The $T^a$ are generators of the Lie algebra of the group $G$.

In the presence of $D$-branes, the string worldsheet has boundaries
and we have to impose an appropriate gluing condition for the currents.
The most natural choice is the Neumann-type
\footnote{Here we take the closed-string point of view;
the boundary is placed at $\text{Re}\, z = 0$. 
In the open-string channel, we take the boundary at $\text{Im}\, z = 0$
and the Neumann boundary condition turns out to be
$J^a_n = \Jt^a_{-n}$.
}
\beq
\label{glue}
  J^a_n + \Jt^a_{-n} = 0 \, ,
\eeq
which preserves the diagonal part of the $G \times G$ symmetry
on the worldsheet.
Since the energy-momentum tensor 
$T(z) = \sum_n L_n z^{-n-2}$ of the WZW model is 
a quadratic form of the current $J(z)$, 
the above condition assures that half of the conformal symmetry
is also preserved:
\beq
  L_n - \Lt_{-n} = 0 \, .
\eeq
Hence, the gluing condition \eqref{glue} is compatible with
the conformal invariance. 

As is shown in \cite{AS}, the Neumann-type condition \eqref{glue}
implies that the boundary of 
the worldsheet is constrained within one of the conjugacy classes 
of $G$,
\beq
  g(z,\zbar)|_{z+\zbar=0} \in \class(t) \, .
  \label{boundary_condition}
\eeq
In other words, the worldvolume of the $D$-brane coincides with
the conjugacy classes. 
In order to understand this result, 
let us consider the wave function $\psi(x)$ for the string zero mode. 
As eq.\eqref{LR} shows,
$J + \Jt$ is the generator of the adjoint action
$x \rightarrow g x g^{-1}$.
The gluing condition \eqref{glue} therefore means that 
the wave function
should obey the relation $\psi(g x g^{-1}) = \psi(x)$,
i.e., the wave function is a class function. 
Among the class functions, we have 
the $\delta$-function $\delta_t$ given in \eqref{delta}
that is concentrated on a specific conjugacy class $\class(t)$.
If we take it as the wave function of string, the corresponding state
is nothing but a $D$-brane wrapped around the conjugacy class 
$\class(t)$.

For $G = SU(2) \cong S^3$, a generic conjugacy class
is isomorphic to a 2-sphere $S^2$.
In addition to this, there are two conjugacy classes,
$\class(1)$ and $\class(-1)$, consisting of a point.
We can parametrize the conjugacy classes by the element
$t = e^{i \theta \sigma^3}$ of the maximal torus $T \cong S^1$.
\footnote{
Our convention is
$\sigma^1 = 
\bigl( \begin{smallmatrix} 0 & 1 \\ 1 & 0 \end{smallmatrix} \bigr), \,
\sigma^2 = 
\bigl( \begin{smallmatrix} 0 & -i \\ i & 0 \end{smallmatrix} \bigr), \,
\sigma^3 = 
\bigl( \begin{smallmatrix} 1 & 0 \\ 0 & -1 \end{smallmatrix} \bigr)$.
}
More precisely, what parametrizes the conjugacy class is
$T/W$, where $W$ is the Weyl group of $G$.
For $G = SU(2)$, $W = \Z_2$ and $t$ takes values in $S^1/\Z_2$. 
The parameter $\theta \in [0, \pi]$ 
coincides with the latitude in $S^3$, where
$\theta=0$ and $\pi$ correspond to the north and the south pole 
of $S^3$, respectively.
From the quantization of the Dirac flux on the worldvolume,
only a finite number of classes are allowed \cite{AS,G},
namely 
$\theta = \pi \frac{2\alpha}{k},\, \alpha = 0, 1/2, 1, \cdots, k/2$.
We therefore have $k-1$ 2-branes, 
which correspond to the `regular' conjugacy classes,
and two 0-branes, which are located at the north and the south pole. 

If we suppose that the worldvolume of the $D$-brane exactly 
coincides with the conjugacy class, the wave function of the string
zero mode for the boundary state would be the $\delta$-function
\eqref{delta}.
For the $D$-brane at $\theta = \pi \frac{2\alpha}{k}$, 
we obtain the wave function $\psi^k_\alpha$ as follows
\beq
  \psi^k_\alpha(\theta) \equiv
    \delta_{t = e^{i \pi \frac{2\alpha}{k} \sigma^3}}
	(e^{i \theta \sigma^3})
  = \sum_{j = 0, 1/2, 1, \cdots} \!\!
    \chi_j(\pi \tfrac{2\alpha}{k}) \, \overline{\chi_j(\theta)} \, .
	\label{classical_wave_function}
\eeq
Here, we write the $su(2)$ character $\chi_j$ 
of the irreducible representation with spin $j$
as a function of $\theta$
rather than the group element $t$,
\beq
  \chi_j(\theta) = \trace{V_j} e^{i \theta \sigma^3}
    = \frac{\sin((2j+1)\theta)}{\sin \theta} \, ,
	\label{su2_character}
\eeq
where $V_j$ is the representation space of spin $j$.

\vspace{.5\baselineskip}
We can analyze the branes in $SU(2)$ from the point of view of
the exact CFT.
The WZW model for $G = SU(2)$ with level $k$ is described by
a CFT with the current algebra $\widehat{su}(2)_k$ and
the diagonal modular invariant partition function.
In this CFT, there are a finite number of primary fields,
which are labeled by the $\widehat{su}(2)_k$ integrable highest weight 
$j = 0, 1/2, 1, \cdots, k/2$.
The Hilbert space of the CFT therefore takes the form
\beq
  \bigoplus_{j=0, 1/2, 1, \cdots, k/2} 
  \widehat{V}_j \otimes \widehat{V}_{\overline{\jmath}} \, ,
\eeq
where $\widehat{V}_j$ is the irreducible representation
of $\widehat{su}(2)_k$ with spin $j$ and $\overline{\jmath}$ is
the complex conjugate of the $su(2)$ representation $j$.
The left-moving sector of string acts on $\widehat{V}_j$ while
the right-moving one acts on $\widehat{V}_{\overline{\jmath}}$.
This structure of the Hilbert space is analogous to 
the spectrum \eqref{L2G} of a particle on $G$.
Actually, the ground states of 
$\widehat{V}_j$ transform as the representation of spin $j$
under the action of the $su(2)$ subalgebra $J^a_0$.
They represent the wave function of the zero modes of string and
we denote them as $V_j$.
Schematically, 
$\widehat{V}_j = \{J^a_{-n} \} V_j$. 

For the worldsheet with boundaries, 
we have to impose the gluing condition \eqref{glue}
at the boundary. 
The boundary states subject to the gluing condition \eqref{glue}
have been obtained by Ishibashi \cite{Ishibashi} and are called 
the Ishibashi states. 
They are labeled by the spin $j$ and we denote them as $\Ishibashi{j}$
\beq
  \Ishibashi{j} = \sum_N \ket{j, N} \otimes \overline{\ket{j, N}} \, ,
  \label{Ishibashi}
\eeq
where $\ket{j, N}$ is an orthonormal basis of $\widehat{V}_j$
and $\overline{\ket{j, N}}$ is its complex conjugate.
One can show that
the Ishibashi states $\Ishibashi{j}$ satisfy the following equation
\beq
  {}_I\!\dbra{j} 
  q^{\frac{1}{2}(L_0 + \Lt_0 - \frac{c}{24})} 
  z^{\frac{1}{2}(J^3_0 - \Jt^3_0)}
  \Ishibashi{j'} = \delta_{jj'} \hat{\chi}_j(\tau, \nu) \, ,
  \label{annulus_Ishibashi}
\eeq
We write $q = e^{2 \pi i \tau}$, $z = e^{2 \pi i \nu}$,
and the central charge $c$ takes the value $c = \frac{3k}{k+2}$. 
$\hat{\chi}_j$ is the character of the irreducible representation
$\widehat{V}_j$ of $\widehat{su}(2)_k$,
\beq
  \hat{\chi}_j(\tau, \nu) = 
  \trace{\hat{V}_j} q^{L_0 - \frac{c}{24}} z^{J^3_0} 
  = \frac{\vartheta_{2j+1,k+2} - \vartheta_{-2j-1,k+2}}
     {\vartheta_{1,2} - \vartheta_{-1,2}} (\tau, \nu)\, ,
	 \label{affine_character}
\eeq
where the theta function $\vartheta_{a,b}$ is 
\beq
  \vartheta_{a,b}(\tau,\nu) = 
  \sum_{m \in \Z + \frac{a}{2b}} \exp(2\pi i b(\tau m^2 + \nu m)) \, .
\eeq

A generic boundary state satisfying the gluing condition 
\eqref{glue} is a linear combination of the Ishibashi states.
In this sense, the Ishibashi state is
the building block of the boundary states.
However, we can not take an arbitrary combination of them
because of the constraint from the modular invariance \cite{Cardy}.
Suppose that we have two boundary conditions
$\alpha$ and $\beta$ satisfying the gluing condition \eqref{glue},
and denote the corresponding boundary
state as $\dket{\alpha}$ and $\dket{\beta}$.
We can calculate the annulus amplitude with the boundary conditions
$\alpha$ and $\beta$ in two different ways.
From the closed string channel,
the amplitude can be calculated as 
\beq
  \dbra{\alpha} 
  q^{\frac{1}{2}(L_0 + \Lt_0 - \frac{c}{24})} 
  z^{\frac{1}{2}(J^3_0 - \Jt^3_0)}
  \dket{\beta} \, .
  \label{annulus}
\eeq
From the open string channel, the above
amplitude is the trace over the open string Hilbert space and
counts the spectrum of the open string with the boundary conditions
$\alpha$ and $\beta$.
Since the gluing condition \eqref{glue} preserves the
$\widehat{su}(2)_k$ symmetry, 
the spectrum is a direct sum of the irreducible
representations of $\widehat{su}(2)_k$. 
The annulus amplitude \eqref{annulus} should therefore be a sum of
the $\widehat{su}(2)_k$ characters with integer coefficients.
This requirement for the annulus amplitude restricts the form 
of the boundary states. Cardy \cite{Cardy} solved this problem
and obtained the allowed form of the states. 
We call them Cardy's states and denote them as $\Cardy{\alpha}$
\beq
  \Cardy{\alpha} = 
  \sum_{j} S_{\alpha j} \frac{1}{\sqrt{S_{0j}}} \Ishibashi{j} \, .
  \label{Cardy}
\eeq
Here the sum is over all the integrable highest weights 
$j = 0, 1/2, 1, \cdots, k/2$.
$S_{i j}$ is the modular transformation matrix
\footnote{Here, we set $\nu = 0$.}
\beq
  \hat{\chi}_i(-1/\tau) = \sum_j S_{ij} \hat{\chi}_j(\tau) \, .
\eeq
For $\widehat{su}(2)_k$, it takes
the following form
\beq
  S_{ij} = 
  \sqrt{\frac{2}{k+2}} 
  \sin\!\left(\frac{\pi (2i+1)(2j+1)}{k+2}\right) \, .
\eeq
As is seen from the definition \eqref{Cardy}, Cardy's states
$\Cardy{\alpha}$ are also labeled by the integrable highest weight
$\alpha = 0, 1/2, 1, \cdots, k/2$. 
Hence we have $k+1$ boundary states, 
or $k+1$ branes, which are subject to the gluing
condition \eqref{glue} and satisfy the modular invariance.

The map \eqref{map} between the square integrable function and
the state of quantum mechanical particle on $G$
enables us to read off the wave function corresponding to
the boundary state.
For example, the wave function for the Ishibashi state 
\eqref{Ishibashi} can be obtained as follows
\beq
  \begin{split}
  \Ishibashi{j} &= \sum_{-j \le m \le j} 
  \ket{j, m} \otimes \overline{\ket{j, m}}
  + \text{states with oscillators} \\
  & \quad\quad\quad\quad \longmapsto \quad
      \sqrt{2j + 1} \!\sum_{-j \le m \le j}\! \overline{\pi}_{mm}
	= \sqrt{2j + 1} \, \overline{\chi}_j \, .
  \end{split}
\eeq
Here $\ket{j, m}$ is the orthonormal basis of $V_j$ 
and $m$ is the weight $J^3_0 = m$. 
This equation shows that the Ishibashi state $\Ishibashi{j}$ 
corresponds to the $su(2)$ character $\chi_j$ up to normalization.
Using this fact, we can obtain the wave function 
$\widehat{\psi}^k_\alpha$ corresponding to Cardy's state
$\Cardy{\alpha}$
\footnote{%
The wave function for Cardy's states is also discussed in 
ref.\cite{FFFS}. 
}
\beq
  \begin{split}
  \Cardy{\alpha} &= 
  \sum_{j=0, 1/2, 1, \cdots, k/2} \! 
  S_{\alpha j} \frac{1}{\sqrt{S_{0j}}} \Ishibashi{j} \\
  & \quad\quad \longmapsto \quad
  \widehat{\psi}^k_\alpha(\theta) \equiv
  \frac{S_{\alpha 0}}{\sqrt{S_{00}}} \sum_{j=0, 1/2, 1, \cdots, k/2}
  \sqrt{\frac{2j+1}{[2j+1]_q}} 
  \chi_j(\pi \tfrac{2\alpha + 1}{k+2}) \overline{\chi}_j(\theta) \, .
  \end{split}
  \label{quantum_wave_function}
\eeq
Here we have used the relations
\bea
  \frac{S_{\alpha j}}{S_{\alpha 0}} &= 
     \chi_j(\pi \tfrac{2\alpha + 1}{k+2}) \, , \\
  \frac{S_{0j}}{S_{00}} &= \chi_j(0) = 
  [2j+1]_q \, , \quad q = e^{i\frac{2\pi}{k+2}} \, ,
\eea
and the $q$-number $[n]_q$ defined by
\beq
  [n]_q = 
  \frac{q^{\frac{n}{2}} - q^{-\frac{n}{2}}}
       {q^{\frac{1}{2}} - q^{-\frac{1}{2}}} \, .
\eeq
One can see that the wave function $\widehat{\psi}^k_\alpha$
for Cardy's state $\Cardy{\alpha}$ 
has almost the same form as the wave function $\psi^k_\alpha$
\eqref{classical_wave_function}
obtained under the assumption that the worldvolume of the $D$-brane
exactly coincides with the conjugacy class.
The differences between these two are:
(1) $\widehat{\psi}^k_\alpha$ has the additional factor
$\sqrt{\frac{2j+1}{[2j+1]_q}}$ 
(2) The sum over the spin $j$ is restricted to $j \le k/2$ for 
$\widehat{\psi}^k_\alpha$
(3) $k$ and $2\alpha$ is shifted to $k+2$ and $2\alpha+1$,
respectively, in $\widehat{\psi}^k_\alpha$.
Clearly, these differences disappear in the limit of
$k \rightarrow \infty$ with $\frac{2\alpha}{k}$ fixed,
since $[n]_q$ reduces to $n$ as
$q \rightarrow 1$. 
Therefore, we can regard the wave function $\psi^k_\alpha$ 
as the $k \rightarrow \infty$ (or the classical) limit of 
the wave function $\widehat{\psi}^k_\alpha$ for Cardy's state.
For finite $k$, the $D$-brane wrapped
around the conjugacy class is described by Cardy's state.
We can consider that the difference between $\psi^k_\alpha$ and
$\widehat{\psi}^k_\alpha$ for finite $k$ is the quantum effect,
which vanish in the classical limit $k \rightarrow \infty$.

\vspace{.5\baselineskip}
Instead of the gluing condition \eqref{glue}, one can take 
a `twisted' form \cite{KO}
\beq
  J^a_n + \omega(\Jt^a_{-n}) = 0 \, ,
  \label{glue_twisted}
\eeq
where $\omega$ is a Lie algebra automorphism. 
This includes the Dirichlet boundary condition
in the case of the flat background, 
for which $\omega$ acts as a rotation
of the right-moving coordinates.
Since $\omega$ does not change the energy-momentum tensor,
the conformal invariance is also preserved under the twisted gluing 
condition \eqref{glue_twisted}. 
The construction of the boundary states proceeds in the way
similar to the ordinary gluing condition  \cite{KO}.

For the WZW model, a natural choice for $\omega$ is an automorphism
of the horizontal subalgebra. 
For such $\omega$, 
the worldvolume corresponding to the gluing condition
\eqref{glue_twisted} is a `twisted' version of
the conjugacy class \cite{FFFS}
\beq
  \class^\omega(t) = \{g t \omega(g)^{-1}\, , \, g \in G\} \, .
  \label{class_twisted}
\eeq
This fact is easily understood in terms of the wave function,
in the same way as the Neumann-type condition \eqref{glue}.
Namely, with the gluing condition \eqref{glue_twisted}, 
one can show that the wave function $\psi(x)$ should satisfy 
$\psi(g x \omega(g)^{-1}) = \psi(x)$, which is obviously related
with the twisted conjugacy class \eqref{class_twisted}.

If we take $\omega$ as an inner automorphism 
$\omega(\lambda) = \omega \lambda \omega^{-1}$,
$C^\omega$ is isomorphic to the ordinary conjugacy class \eqref{class}
\beq
  C^\omega(t) = \{g t (\omega g \omega^{-1})^{-1} \, , \, g \in G \}
         = \{g (t \omega) g^{-1} \cdot \omega^{-1}\, , \, g \in G \}
		 = C(t\omega) \omega^{-1} \, .
\eeq
On the other hand, for an outer automorphism, $C^\omega$ has 
in general a topology different from the ordinary class \cite{FFFS}.
One can regard this fact as a generalization of the well-known
behavior of $D$-branes on a torus under the $T$-duality transformation.
As an example, we take a $T$-duality transformation
$X_L(z) + X_R(\zbar) \rightarrow X_L(z) - X_R(\zbar)$, which acts
on the current $\Jt = i\delbar X_R$ as an outer automorphism 
$\Jt \rightarrow - \Jt$. 
The dimension of the $D$-brane worldvolume changes by one under
this transformation, which shows that an outer automorphism can change
the topology, the dimension in this case, of the worldvolume.

\section{Free field realization of the boundary condition}
\subsection{Matching condition}

One of the characteristic features of conformal field theory in
two dimensions is the existence of chiral and anti-chiral
sectors commuting with each other. 
For a string coordinate $X$ of toroidal compactification, 
which takes values in $S^1 \cong \R/\Z$, 
we have a decomposition
\beq
  X(z,\zbar) = X_L(z) + X_R(\zbar) \, .
\eeq
This follows from the equation of motion
$\del \delbar X = 0$. 
Since there is no real holomorphic function except for a constant,
we can not decompose $X$ into $X_L$ and $X_R$ within
the real functions.
This would be the case if we took the Minkowski metric,
instead of the Euclidean metric, on the worldsheet.

One can regard the boundary condition of a string 
as a matching condition
of the left- and the right-moving sectors. 
Actually, the Dirichlet boundary condition
\beq
  X = X_L + X_R = a \,\, \mbox{(constant)}
\eeq
relates the left- and the right-moving coordinates 
at the boundary and can be considered as a matching
condition of $X_L$ and $X_R$. 
Since the Dirichlet condition is mapped to the Neumann condition via
the $T$-duality transformation $X_R \rightarrow -X_R$,
we can also regard the Neumann condition as a matching condition
of the two chiral sectors.
To summarize, we have two types of matching conditions for a single
coordinate $X$:
\begin{subequations}
\begin{alignat}{2}
  X_L &= X_R + a    & \quad& \mbox{(Neumann)}   \, , 
  \label{match_N}\\
  X_L &= - X_R + a  &      & \mbox{(Dirichlet)} \, . 
\end{alignat}
\end{subequations}
From these matching conditions, one can obtain the gluing condition
for the currents $J = i\del X_L$ and $\Jt = i\delbar X_R$.
\footnote{%
In this paper,
we use the word `matching' for the boundary condition including
the chiral zero modes, while the word `gluing' is used for
the condition on the currents, i.e., without the zero modes.
}
To see this,
suppose a holomorphic function $f(z)$ and 
an anti-holomorphic one $\bar{f}(\zbar)$
satisfying the matching condition
$f(z) = \bar{f}(\zbar)$ along the boundary $z + \zbar = 0$.
\footnote{%
Note that we take the closed-string point of view,
in which the boundary of the worldsheet is placed
at $\text{Re}\, z = 0$.
}
Then, along the boundary, their derivatives are related as follows,
\beq
  f'(z) = \frac{d}{dz} f(z)
  = \frac{d}{dz} \bar{f}(-z) 
  = - \bar{f}'(\zbar) \, .
\eeq
Applying this relation to $X_L$ and $X_R$,
the matching condition 
\eqref{match_N} of the Neumann type implies
\beq
  J = i\del X_L = (-i \delbar) (X_R + a) = - i\delbar X_R = -\Jt \, ,
\eeq
which is exactly the Neumann-type gluing condition.

We extend this argument to the case of group manifolds.
In the WZW model on a group $G$,
the $G$-valued field $g(z,\zbar)$ satisfies
the equation of motion
$\del(g^{-1} \delbar g) = 0$, 
which can be solved in the form
\beq
  g(z,\zbar) = g_L(z) g_R(\zbar)^{-1} \, .
  \label{decomposition}
\eeq
In the same way as the case of $S^1$, we should consider that
the fields $g_L$ and $g_R$ take values in the 
complexification $G^{\C}$ of $G$.
If we take the Minkowski metric on the worldsheet, it is possible
to decompose $g$ within $G$. 
Note that this decomposition is determined up to a constant group
element. In fact, $g$ is invariant under the substitution
$g_L \rightarrow g_L M$, $g_R \rightarrow g_R M$, $M \in G$.
Using the above decomposition \eqref{decomposition} of $g$ in
the expressions \eqref{J} for the currents,
one obtains
\beq
\begin{aligned}
  J   &= - k \del g_L \, g_L^{-1} \, , \\
  \Jt &= - k \delbar g_R \, g_R^{-1} \, . 
\end{aligned}
\eeq

As is explained in the previous section, 
the gluing condition \eqref{glue} of the Neumann type corresponds to 
a $D$-brane on the conjugacy class.
More precisely, the gluing condition restricts the form of the wave
function to the class function. Since the gluing condition is linear
on the wave function, we can take any superposition of 
the class functions as the wave function. 
Consequently, the position of $D$-brane is not completely
fixed by the gluing condition.
In order to specify the position of $D$-brane, we have to treat
directly the boundary condition including the zero modes,
i.e., the matching condition,
instead of the gluing condition of the currents. 

We have seen that the boundary condition in the case of $S^1$
can be rewritten as
a matching condition of the left- and the right-moving sectors
including the chiral zero modes.
The boundary condition for the WZW model also admits
the formulation as a matching condition. 
Our proposal is as follows:
\beq
  g_L = g_R t \, , \quad t \in G \, ,
  \quad \text{along the boundary $z+\zbar=0$}.
  \label{match}
\eeq
Because of the ambiguity in the decomposition 
\eqref{decomposition} of $g$,
$t$ is determined up to a conjugation by a group element.
In fact, $t$ transforms as $t \rightarrow M^{-1} t M$,
under the change of variables
$g_L \rightarrow g_L M, g_R \rightarrow g_R M, M \in G$.
Therefore, we can specify only the class $\class(t)$, 
not a single element $t$, in the matching condition \eqref{match}. 
One can confirm that this condition reproduces the boundary 
condition $g|_{z+\zbar=0} \in \class(t)$ 
given in \eqref{boundary_condition}.
Actually,
\beq
  g = g_L g_R^{-1} = g_R t g_R^{-1} \in \class(t) \, .
\eeq
The gluing condition \eqref{glue} of the currents 
also follows from the matching condition \eqref{match}:
\beq
  J = -k \del g_L \, g_L^{-1} 
    = -k (-\delbar) (g_R t) t^{-1} g_R^{-1}
    = k \delbar g_R \, g_R^{-1} = -\Jt \, .
\eeq

We can also write the twisted boundary condition 
$g|_{z+\zbar=0} \in \class^\omega(t)$ in the same way:
\beq
  \omega(g_L) = g_R t \, , \quad t \in G\, ,
  \quad \text{along the boundary $z+\zbar=0$}.
  \label{match_twisted}
\eeq
For the group element $g$, we obtain
\beq
  g = g_L g_R^{-1} = g_L t \omega(g_L)^{-1} \in \class^\omega(t) \, .
\eeq
The currents satisfy the following equation
\beq
  \Jt = -k \delbar g_R \, g_R^{-1} 
      = k\, \omega (\del g_L \, g_L^{-1}) = \omega(-J) \, .
\eeq
Since $J$ can be written as 
$\sum_a J^a T^a = \sum_a J^{\omega(a)} \omega(T^a)$, 
the above equation implies
\beq
  \Jt^{\omega(a)} = - J^a \, ,
\eeq
which is exactly the twisted gluing condition \eqref{glue_twisted}.

\subsection{Free field realization}
We have seen that a Neumann-type boundary condition
can be derived from 
the matching condition \eqref{match} of $g_L$ and $g_R$.
We use this fact to obtain the boundary states for the $D$-brane
wrapped around the conjugacy class of $SU(2)$.

In order to write the boundary states for
the matching condition \eqref{match_twisted} explicitly,
we adopt here the free field realization of $SL(2, \R)$
\cite{GMOM,G2,BF}
\beq
  g =
  \begin{pmatrix} 1         & \gamma \\ 
                  0         & 1        \end{pmatrix}
  \begin{pmatrix} e^{-\phi} & 0 \\ 
                  0         & e^{\phi} \end{pmatrix}
  \begin{pmatrix} 1         & 0 \\ 
                  \gammabar & 1        \end{pmatrix}  \, .
  \label{Gauss}
\eeq
Note that $\gamma, \gammabar \text{ and } \phi$ are real fields
for $g \in SL(2, \R)$. 
In terms of these fields, the worldsheet action $S$ 
for the $SL(2,\R)$ WZW model with level $\tilde{k}$
can be written as
\beq
  S = \frac{1}{4\pi} \int\! d^2\! z
  \left( \del \varphi \delbar \varphi 
         + \beta \delbar \gamma + \betabar \del \gammabar
		 - \beta\betabar e^{-2\varphi/\tilde{\alpha}_+}
		 - \frac{2}{\tilde{\alpha}_+}\varphi \sqrt{g} R \right) \, ,
\eeq
where $\beta$ and $\betabar$ are auxiliary fields conjugate to 
$\gamma$ and $\gammabar$, respectively. 
$\varphi$ is related with $\phi$ by
\beq
  \varphi = \tilde{\alpha}_+ \phi \, , \quad
  \tilde{\alpha}_+ = \sqrt{2 \tilde{k} - 4} \, .
  \label{alpha_tilde}
\eeq
For large $\phi$, 
one can treat the screening charge perturbatively, and
it is possible to consider $(\beta, \gamma)$ and 
$(\betabar, \gammabar)$ as free fields.
We therefore have two pairs of bosonic first-order systems 
and a free boson:
\begin{alignat}{2}
    \gamma(z) &= \sum_n \gamma_n z^{-n}  \, ,   & \quad\quad
    \gammabar(\zbar) &= \sum_n \gammabar_n \zbar^{-n}  \, ,  \notag \\
    \beta(z)  &= \sum_n \beta_n z^{-n-1} \, ,   &
    \betabar(\zbar)  &= \sum_n \betabar_n \zbar^{-n-1} \, ,  \notag
\end{alignat}
\begin{align}
    \varphi(z,\zbar) &= \varphi_L(z) + \varphi_R(\zbar) \, , \\
    \varphi_L(z)     &= x_L - i p_L \ln z
	+i \sum_{n \ne 0} \dfrac{1}{n} \alpha_n z^{-n} \, ,  \notag \\
    \varphi_R(\zbar) &= x_R - i p_R \ln \zbar
	+i \sum_{n \ne 0} \dfrac{1}{n} \alphat_n \zbar^{-n} \, . \notag
\end{align}
These fields satisfy the following OPE,
\beq
\begin{aligned}
  {}& \varphi(z,\zbar) \varphi(w,\wbar) \sim -\ln |z-w|^2 \, , \\
    & \beta(z) \gamma(w) \sim \frac{1}{z-w} \, , \quad
     \betabar(\zbar) \gammabar(\wbar) \sim \frac{1}{\zbar-\wbar} \, .
\end{aligned}
\eeq
The commutation relations among the modes are
\beq
  [\beta_m, \gamma_n] = \delta_{m+n} \, , \quad
  [x_L, p_L] = i \, , \quad
  [\alpha_m, \alpha_n] = m \delta_{m+n} \, .
\eeq
We regard $SU(2)$ with level $k$
as $SL(2,\R)$ with $\tilde{k} = -k$.
We therefore obtain 
\beq
  \varphi = -i \alpha_+ \phi \, , \quad
  \alpha_+ = i \tilde{\alpha}_+ = \sqrt{2 k + 4} \, ,
  \label{quantum_phi}
\eeq
for the case of $SU(2)$ instead of \eqref{alpha_tilde}. 

One can express the currents $J^a$ and $\Jt^a$ as differential 
operators acting on the functions of $(\gamma, \gammabar$, $\phi)$:
\beq
  \begin{aligned}
     J^+ &= -\dfrac{\del}{\del\gamma} \, , \\
	 J^3 &= -\gamma\dfrac{\del}{\del\gamma} 
	         + \dfrac{1}{2}\dfrac{\del}{\del\phi} \, , \\
	 J^- &=  \gamma^2 \dfrac{\del}{\del\gamma}
	         - \gamma\dfrac{\del}{\del\phi}
			 - e^{-2\phi} \dfrac{\del}{\del\gammabar} \, ,
  \end{aligned} \quad\quad
  \begin{aligned}
	 \Jt^+ &= -\gammabar^2 \dfrac{\del}{\del\gammabar}
	         + \gammabar\dfrac{\del}{\del\phi}
			 + e^{-2\phi} \dfrac{\del}{\del\gamma} \, , \\
	 \Jt^3 &=  \gammabar\dfrac{\del}{\del\gammabar} 
	         - \dfrac{1}{2}\dfrac{\del}{\del\phi} \, , \\
     \Jt^- &= \dfrac{\del}{\del\gammabar} \, .
  \end{aligned}
\eeq
In terms of the free fields, these currents can be written as follows
\beq
  \begin{aligned}
     J^+ &= -\beta \, ,\\
	 J^3 &= -\beta\gamma
	         + \dfrac{1}{2}\alpha_+ i\del\varphi \, ,\\
	 J^- &= \beta \gamma^2 
	         - \gamma \alpha_+ i \del\varphi
			 - k \del \gamma \, ,
  \end{aligned} \quad\quad
  \begin{aligned}
	 \Jt^+ &= -\betabar \gammabar^2 
	         + \gammabar \alpha_+ i \delbar \varphi
			 + k \delbar \gammabar \, , \\
	 \Jt^3 &= \betabar \gammabar
	         - \dfrac{1}{2} \alpha_+ i \delbar \varphi \, , \\
     \Jt^- &= \betabar \, .
  \end{aligned}
  \label{Wakimoto}
\eeq
The Sugawara energy-momentum tensor takes the form
\beq
  T = \frac{1}{k+2} J^a J^a
    = \beta \del \gamma 
	  - \frac{1}{2}(\del \varphi)^2 
	  - \frac{1}{\alpha_+} i \del^2 \varphi \, .
  \label{Sugawara}
\eeq
The vertex operator $V_{j, m}$, which has the $su(2)$ spin $j$
and $J^3_0 = m$, can be written as
\beq
  V_{j, m} = \gamma^{j-m} e^{i\frac{2j}{\alpha_+}\varphi} \, .
  \label{vertex}
\eeq
This operator has the conformal dimension 
$\Delta_j = \frac{j(j+1)}{k+2}$.
Since we have a non-vanishing 2-point function
$\cor{V_{-j-1, -m}(z) V_{j,m}(0)} \ne 0$,
the operator $V_{-j-1,-m}$ can be considered as 
the dual of $V_{j, m}$.

From the expression \eqref{Gauss} for $g$,
we can read off the left- and the right-moving part of $g$ as follows
\beq
  \begin{aligned}
  g_L &=
  \begin{pmatrix} 1         & \gamma \\ 
                  0         & 1        \end{pmatrix}
  \begin{pmatrix} e^{-\phi_L} & 0 \\ 
                  0         & e^{\phi_L} \end{pmatrix} \, , \\
  g_R &=
  \begin{pmatrix} 1         & 0 \\ 
                 -\gammabar & 1        \end{pmatrix}
  \begin{pmatrix} e^{\phi_R} & 0 \\ 
                  0         & e^{-\phi_R} \end{pmatrix} \, ,
  \end{aligned}
  \label{gl_gr}
\eeq
where $\phi_L$ and $\phi_R$ are the left- and the right-moving
part of $\phi$, respectively.
The left-moving part $g_L$ is an upper triangular
matrix while $g_R$ is a lower triangular one. 
In other words, $g_L$ ($g_R$) is an element of the Borel group 
$B_+$ ($B_-$)
\footnote{
Here, the Borel group $B_+$ is a group of all the upper triangular
matrices in $SL(2, \C)$.}.
Since $B_+$ overlaps with $B_-$ only at the diagonal matrices,
the matching condition $g_L = g_R t$ is not suitable for
the free field expression \eqref{gl_gr}.
In order to find an appropriate condition, we regard
$SU(2)$ as a subgroup of $SL(2, \C)$.
This is also
necessary to perform an analytic continuation from $SU(2)$
to $SL(2, \R)$, for which the free field realization \eqref{Gauss}
is introduced. 
In $SL(2, \C)$, we have a non-trivial involution
$g \rightarrow (g^\dagger)^{-1}$, which reduces to
a trivial one for $SU(2)$. 
Correspondingly, we have two types of extensions of 
the condition $g_L = g_R t$ to $SL(2, \C)$:
\begin{subequations}
\begin{align}
  g_L                &= g_R t \, , \label{match_original} \\
  (g_L^\dagger)^{-1} &= g_R t \, . \label{match_modify}
\end{align}
\end{subequations}
If we stay in $SU(2)$, these two conditions
are identical.
However, $(g^\dagger)^{-1} \ne g$ for $g \in SL(2, \C)$
and the difference between these two expressions is meaningful.
Here we take the latter choice \eqref{match_modify} to describe
the $D$-brane in $SU(2)$. 
Since the involution $g \rightarrow (g^\dagger)^{-1}$ maps 
$B_+$ to $B_-$, the condition \eqref{match_modify} can be applied to
the free field expression \eqref{gl_gr}.

As noticed above, 
we take the free fields $(\gamma, \gammabar, \phi)$ to be real.
Hence, the matching condition \eqref{match_modify} can be 
written as follows
\beq
  \begin{pmatrix} 1      & 0 \\ 
                 -\gamma & 1              \end{pmatrix}
  \begin{pmatrix} e^{\phi_L} & 0 \\ 
                  0          & e^{-\phi_L} \end{pmatrix} =
  \begin{pmatrix} 1         & 0 \\ 
                 -\gammabar & 1           \end{pmatrix}
  \begin{pmatrix} e^{\phi_R + i a} & 0 \\ 
                  0                & e^{-\phi_R - i a} \end{pmatrix} 
  \, .
\eeq
Here we write the element $t \in T/W$, which represents 
the conjugacy class, as
$t = \bigl( 
     \begin{smallmatrix} e^{i a} & 0 \\ 
                         0       & e^{- i a}
	 \end{smallmatrix} \bigr) $, $a \in [0, \pi]$. 
From this expression, the boundary condition for the free fields
follows immediately 
\beq
  \begin{array}{rcl}
  \gamma &=& \gammabar \, , \\[\jot]
  \varphi_L &=& \varphi_R + \alpha_+ a \, ,
  \end{array}
\eeq
where we used the relation \eqref{quantum_phi} to rewrite $\phi$
with $\varphi$. 
One can see that $\varphi$ satisfies the Neumann-type matching
condition \eqref{match_N}. 
In terms of modes, this condition can be written as
\beq
  \begin{aligned}
  \gamma_n - \gammabar_{-n} &= 0 \, , \quad \quad &
  x_L - x_R &=  \alpha_+ a \, , \\
  \beta_n + \betabar_{-n}   &= 0 \, , &
  \alpha_n + \alphat_{-n}   &= -\dfrac{2}{\alpha_+} \delta_{n,0} \, ,
  \end{aligned}
  \label{boundary}
\eeq
where we denote $p_L$ and $p_R$ as $\alpha_0$ and $\alphat_0$, 
respectively.
The condition for $\beta$ and $\betabar$ follows from that for 
$\gamma$ and $\gammabar$. 

The condition for $\alpha_0$ and $\alphat_{0}$
differs from the usual Neumann-type.
This is because, as is seen from \eqref{Sugawara},
$\varphi$ has a background charge. 
To understand this, let us consider a current
($W(z)$, $\widetilde{W}(\zbar)$)
with background charge $Q$,
which has the following OPE with
the energy-momentum tensor $T(z)$:
\beq
  T(z) W(w) \sim \frac{1}{(z-w)^3} (-Q) + \frac{1}{(z-w)^2} W(w)
                 + \frac{1}{z-w} \del W(w) \, .
  \label{anomalous_OPE}
\eeq
In terms of modes, we have
\beq
  [L_m, W_n] = -n W_{m+n} - \frac{Q}{2} m(m+1) \delta_{m+n} \, .
\eeq
Suppose that $W(z)$ and its anti-holomorphic counterpart
$\widetilde{W}(\zbar)$
satisfy the gluing condition
\beq
  W_n + \widetilde{W}_{-n} = 0 \, , \quad n \ne 0 \, .
\eeq
Assuming that the boundary condition preserves 
the conformal invariance, one can show
\beq
  0 = [L_m - \Lt_{-m}, W_{-m} + \widetilde{W}_{m}]
    = m(W_0 + \widetilde{W}_0 -Q)  \, ,
\eeq
which means that the boundary condition consistent with the OPE
\eqref{anomalous_OPE} should be
\beq
  W_n + \widetilde{W}_{-n} = Q \delta_{n,0} \, .
  \label{anomaly}
\eeq
Substituting ($i\del \varphi_L$, $i\delbar \varphi_R$) for
($W, \widetilde{W}$), we obtain the boundary condition \eqref{boundary}
for $\alpha_n$ and $\alphat_{-n}$. 

The involution $g \rightarrow (g^\dagger)^{-1}$ acts on the element
$T$ of the horizontal Lie algebra as $T \rightarrow -T^\dagger$.
Hence the current $J^a$ is mapped as follows:
$(J^+, J^3, J^-) \rightarrow (-J^-, -J^3, -J^+)$.
Accordingly, from the matching condition \eqref{match_modify},
we obtain the gluing condition
\beq
  \begin{aligned}
     J^+_n - \Jt^-_{-n}  &= 0 \, , \\
	 J^3_n - \Jt^3_{-n}  &= 0 \, , \\
	 J^-_n - \Jt^+_{-n}  &= 0 \, .
  \end{aligned}
\eeq
Note that
$(-\Jt^-, -\Jt^3, -\Jt^+)$ have exactly the same form as 
$(J^+, J^3, J^-)$
if we substitute $(\beta, \gamma, \varphi_L)$
for $(\betabar, \gammabar, \varphi_R)$ in \eqref{Wakimoto}.
The above gluing condition therefore follows immediately
from the boundary condition \eqref{boundary}.

\subsection{Boundary states}
Let us now describe the construction of the boundary states satisfying 
the boundary condition \eqref{boundary}. 
We treat $(\beta, \gamma)$ and $\varphi$ separately, 
and put them together after completion of the states in each sector. 

We begin with the $(\beta, \gamma)$-sector. 
The boundary condition takes the form
\beq
  \begin{aligned}
  \gamma_n - \gammabar_{-n} &= 0 \, , \\
  \beta_n + \betabar_{-n}   &= 0 \, .
  \end{aligned}
  \label{boundary_bg}
\eeq
This condition preserves the ghost number current
$(J_{\beta\gamma}, \Jt_{\beta\gamma})
 = (\beta\gamma, \betabar \gammabar)$ except for the zero mode.
Since the ghost number current has anomaly in the OPE with 
the energy-momentum tensor, we can apply the previous argument,
in particular \eqref{anomaly}, to obtain 
the relation for the ghost numbers
\beq
  Q_{\beta\gamma} + \widetilde{Q}_{\beta\gamma} = -1 \, ,
  \label{ghost_number}
\eeq
where $(Q_{\beta\gamma}, \widetilde{Q}_{\beta\gamma}) 
 = (\oint\! J_{\beta\gamma}, \oint\! \Jt_{\beta\gamma})$.

As is well-known, the ground state of the $(\beta, \gamma)$-system
is labeled by the bosonic sea-level, which is called a picture.
More specifically, the vacuum $\ket{q}$ in the $q$-picture is 
characterized as follows
\beq
  \begin{aligned}
    \gamma_n \ket{q} &= 0  &\quad\quad  n &\ge 1 + q \, , \\
	\beta_n  \ket{q} &= 0  &\quad\quad  n &\ge - q \, .
  \end{aligned}
\eeq
Note that the vacuum $\ket{q}$ has non-vanishing ghost number, 
namely,
\beq
  Q_{\beta\gamma} \ket{q} = q \ket{q} \, .
\eeq
Hence, the picture of the left-moving sector is related with
that of the right-moving sector through eq.\eqref{ghost_number},
and the ground state takes the form $\ket{q}_L \otimes \ket{-1-q}_R$.
On this ground state, we can construct
the boundary state satisfying the condition \eqref{boundary_bg}
as follows
\beq
  \dket{q}_{\beta\gamma} =
  \exp\left[
    - \sum_{n \ge -q} \gamma_{-n} \betabar_{-n}
	- \sum_{n \ge 1 + q} \beta_{-n} \gammabar_{-n}
	\right]
  \ket{q}_L \otimes \ket{-1-q}_R \, .
  \label{state_bg}
\eeq

Let us consider the $\varphi$-sector. 
The boundary condition is
\beq
  \begin{aligned}
  x_L - x_R &=  \alpha_+ a \, , \\
  \alpha_n + \alphat_{-n}   &= -\dfrac{2}{\alpha_+} \delta_{n,0} \, ,
  \end{aligned}
  \label{boundary_phi}
\eeq
which is almost the Neumann condition except for the zero mode.
We label the ground state of the oscillators $\alpha_n$
by the momentum $p_L = \alpha_0$
\beq
  \begin{aligned}
	\alpha_n  \ket{k} &= 0         & \quad n > 0 \, , \\
    p_L \ket{k} &= k \ket{k}  \, . & 
  \end{aligned}
\eeq
It is convenient to introduce another parametrization of the states
defined by
\beq
  \rket{n} = \ket{\frac{n-1}{\alpha_+}} \, .
\eeq
The momentum $n$ takes integer value 
since $\rket{n}$ corresponds to the vertex operator
$e^{i \frac{n-1}{\alpha_+} \varphi}$ with spin $\frac{n-1}{2}$,
which should be half integer for the unitary irreducible
representation.
In this notation, the boundary condition
$\alpha_0 + \alphat_0 = -\frac{2}{\alpha_+}$ is solved
in the form $\rket{n}_L \otimes \rket{-n}_R$.
One can see, from \eqref{vertex}, that 
the $su(2)$ spin $j$ is related with the momentum $n$ as $n = 2j+1$,
and that $\rket{-n}$ is dual of $\rket{n}$.
The boundary state is obtained by gluing the state with spin 
$j = \frac{n-1}{2}$ and its dual with spin $-j-1$.
Since the condition for the oscillators is of the usual Neumann-type,
we can write 
the boundary state $\dket{n}_\varphi$ in the $\varphi$-sector
as follows
\beq
  \dket{n}_{\varphi} =
  \exp\left[
    - \sum_{m > 0} \frac{1}{m} \alpha_{-m} \alphat_{-m}
	\right]
  \rket{n}_L \otimes \rket{-n}_R \, .
  \label{state_phi}
\eeq
We treat the boundary condition for the zero mode later.

We combine the above results, \eqref{state_bg} and \eqref{state_phi},
to obtain the full boundary state subject to the boundary condition
\eqref{boundary}, namely,
\beq
  \dket{n} = 
    \begin{cases}
	\dket{n}_{\varphi} \otimes \dket{0}_{\beta\gamma} & 
	\text{if $n > 0$} \, , \\
	\dket{n}_{\varphi} \otimes \dket{-1}_{\beta\gamma} & 
	\text{if $n < 0$} \, . 
	\end{cases}
	\label{state_full}
\eeq
Here we take the picture of the $(\beta, \gamma)$-system
as $q=0$ or $-1$ according to the sign of the momentum $n$.
With this choice of the picture, the state with positive spin $j \ge 0$
always appears in the $0$-picture while their dual with spin $-j-1$
does in the $-1$-picture. 
It is helpful to write down the first several terms of
$\dket{n}$ and $\dket{-n}$:
\beq
\begin{aligned}
  \dket{n} &= \rket{n}_L \otimes \rket{-n}_R 
      - \gamma_0 \rket{n}_L \otimes \betabar_0 \rket{-n}_R
  + \frac{1}{2}\gamma_0^2 \rket{n}_L \otimes \betabar_0^2 \rket{-n}_R
  + \cdots \, , \\
  \dket{-n} &= \rket{-n}_L \otimes \rket{n}_R 
      - \beta_0 \rket{-n}_L \otimes \gammabar_0 \rket{n}_R
  + \frac{1}{2}\beta_0^2 \rket{-n}_L \otimes \gammabar_0^2 \rket{n}_R
  + \cdots \, .
\end{aligned}
\eeq
From this expression, one can see that
the left-moving sector of $\dket{n}$ is 
a highest-weight representation of $su(2)$ while 
that of $\dket{-n}$ is a lowest-weight one,
\footnote{
Note that the Fock space of the free fields forms a reducible
representation of $su(2)$.
One needs to take the BRST cohomology in order to obtain
an irreducible representation \cite{BF}.
}
since $\beta_0$ is the raising operator of $su(2)$
(see eq.\eqref{Wakimoto}). 
In other words, the choice of the picture ($0$ or $-1$) 
corresponds to the choice of the representation
(highest-weight or lowest-weight). 

Since the conjugacy class does not change by the action of 
the Weyl group $W$, 
the corresponding boundary state also
should be invariant under the action of $W$.
For $su(2)$, $W = \Z_2$, and 
its non-trivial element acts on the weight $\lambda$
as the reflection $\lambda \rightarrow -\lambda$.
As we show above, 
the left-moving sector of the boundary state $\dket{n}$ 
is a highest-weight representation of $su(2)$.
Hence, the Weyl reflection of $\dket{n}$ has a lowest-weight
representation in the left-moving sector. 
If we write $n = 2j+1$ for $n>0$,
the highest weight of $\dket{n}$ is $j$ while the lowest-weight
of $\dket{-n}$ is $-j$. 
Therefore, the Weyl reflection of $\dket{n}$ is $\dket{-n}$, and
the boundary state $\dket{n}^W$ 
invariant under the Weyl group $W$ takes the following form
\beq
  \dket{n}^W \equiv \dket{n} + \dket{-n} \, .
  \label{state_W}
\eeq
This state $\dket{n}^W$ should be considered as
a building block of the boundary
state for the $D$-brane on the conjugacy class.

We can write the states conjugate to $\dket{n}$ by using the 
hermitian conjugate of the modes:
\bea
  \gamma_n^\dagger &= \gamma_{-n} \, , \notag \\
  \beta_n^\dagger  &= - \beta_{-n} \, , \\
  \alpha_n^\dagger &= - \alpha_{-n} \, . \notag
\eea
Here, $\gamma$ is taken to be real.
We take $\varphi$ to be an anti-hermitian field
since $\varphi$ is related with the real boson $\phi$
through $\varphi = -i \alpha_+ \phi$. 
Therefore the state $\rket{n}$ has
non-vanishing inner product with $\rket{-n}$,
which we normalize as $(-n | n ) = 1$.
Together with the $(\beta, \gamma)$-sector, we obtain
\beq
  \rbra{-n, q= -1} n, q=0) = 1 \, .
\eeq
The explicit form of $\dbra{-n} = \dket{-n}^\dagger$ for $n > 0$ is
\beq
  \begin{split}
  \dbra{-n} &=
  {}_L\rbra{-n, q=-1} \otimes {}_R\rbra{n, q=0} \\
  & \quad\quad \times
  \exp \left[
      \sum_{m \ge 1} \gamma_m \betabar_m
	+ \sum_{m \ge 0} \beta_m \gammabar_m
	- \sum_{m > 0} \frac{1}{m} \alpha_m \alphat_m
	\right] \, .
  \end{split}
  \label{state_conjugate}
\eeq

We can calculate the annulus amplitude \eqref{annulus}
using the boundary state \eqref{state_full} and its conjugate
\eqref{state_conjugate}.
The result is
\beq
  \dbra{-n} 
    q^{\frac{1}{2}(L_0 + \Lt_0 - \frac{c}{24})} 
    z^{\frac{1}{2}(J^3_0 + \Jt^3_0)}
  \dket{n} = 
  \begin{cases}
    \hat{\chi}^0_n(\tau, \nu) & \text{if $n > 0$}\, , \\
	\hat{\chi}^{-1}_n(\tau, \nu) = 
	  -\hat{\chi}^0_n(\tau, \nu) & \text{if $n < 0$}\, .
  \end{cases}
  \label{annulus_free}
\eeq  
Here we introduced the character $\hat{\chi}^q_n$ 
of the free field Fock space $F^q_n$, which is freely generated from 
the ground state $\rket{n} \otimes \ket{q}$ 
by the action of the creation modes.
For the 0-picture, which corresponds to 
the highest-weight representation,
we obtain
\beq
  \hat{\chi}^0_n(\tau, \nu) 
   = \trace{F^0_n} q^{L_0 - \frac{c}{24}} z^{J^3_0} 
   = \frac{q^\frac{n^2}{4(k+2)} z^{\frac{n}{2}}}
     {\vartheta_{1,2}(\tau,\nu) - \vartheta_{-1,2}(\tau,\nu)} \, .
	 \label{free_character}
\eeq
One can also show that
$\hat{\chi}^{-1}_n(\tau, \nu) = - \hat{\chi}^0_n(\tau,\nu)$. 
Hence, the lowest-weight representation ($-1$-picture) behaves
in the character
like the highest-weight representation ($0$-picture) 
with negative norm. 
For the $W$-invariant state $\dket{n}^W$, the annulus amplitude
takes the form
\beq
  \begin{split}
  {}^W\dbra{n} 
    q^{\frac{1}{2}(L_0 + \Lt_0 - \frac{c}{24})} 
    z^{\frac{1}{2}(J^3_0 + \Jt^3_0)}
  \dket{n}^W &= 
  \hat{\chi}^0_n(\tau, \nu) + \hat{\chi}^{-1}_{-n}(\tau, \nu) \\
  &= \hat{\chi}^0_n(\tau, \nu) - \hat{\chi}^0_{-n}(\tau, \nu) \, .
  \end{split}
  \label{annulus_W}
\eeq
This amplitude can be regarded as a regularized inner product
of $\dket{n}^W$ with itself.
One can verify that, for each power of $q$, 
a finite number of states contribute
to the inner product \eqref{annulus_W}.
This follows from the fact that 
the inner product \eqref{annulus_W} is written in the form of 
a difference
of two characters $\hat{\chi}^0_n$ and $\hat{\chi}^0_{-n}$.
Since $\hat{\chi}^0_n$ is the character of the representation
with the highest-weight $\frac{n-1}{2}$, 
the number of states that contributes to the lowest power of $q$
can be calculated as
$\frac{n-1}{2} - \frac{-n-1}{2} = n$.
This number expresses the norm of the wave function corresponding
to the boundary state $\dket{n}^W$. 
The normalized boundary state is therefore
$\frac{1}{\sqrt{n}}\dket{n}^W$. 

In order to complete the boundary state for the conjugacy class,
we have to take an appropriate linear combination 
of the states \eqref{state_W}
to form a $\delta$-function corresponding to the boundary condition 
$x_L - x_R =  \alpha_+ a$ for the zero mode of $\varphi$.
We should consider functions on $S^1/\Z_2$, 
since $a$ parametrizes the position of the conjugacy classes
that takes values in $T/W \cong S^1/\Z_2$.
From Weyl's integral formula, the relevant functions are 
the numerator of the $SU(2)$ character \eqref{su2_character}:
$\sin((2j+1)\theta), \, j = 0, 1/2, 1, \cdots$. 
These functions are orthogonal to each other with respect to
the following inner product
\beq
  \int_0^\pi \! d\theta \sin(m \theta) \sin(n \theta)
  = \frac{\pi}{2} \delta_{m,n} \, .
\eeq
The $\delta$-function for these functions can be written as
\beq
  \delta(\theta, a) =
  \frac{2}{\pi}\sum_{n \ge 1} \sin(n a) \sin(n \theta) \, .
  \label{delta_T}
\eeq
Since the norm of $\sin(n \theta)$ is independent of $n$,
we can identify the wave function $\sin(n \theta)$
with the normalized boundary state $\frac{1}{\sqrt{n}} \dket{n}^W$. 

If we take this $\delta$-function to describe the wave function
of the zero modes, the boundary state for the $D$-brane
wrapped around the conjugacy class at $\theta=a$ can be written as
\beq
  \dket{a}_{\text{brane}} = \sum_{n \ge 1}
  \sin(n a) \frac{1}{\sqrt{n}} \dket{n}^W \, .
  \label{classical_brane}
\eeq
This should be regarded as the boundary state valid in
the $k \rightarrow \infty$ limit, since
the annulus amplitude \eqref{annulus_W} differs from 
the $\widehat{su}(2)$ character $\hat{\chi}_j$ if $k$ is finite.
Actually, one can express $\hat{\chi}_j$ in terms of 
the free field character $\hat{\chi}^0_n$:
\beq
  \hat{\chi}_j = 
    \sum_{l \in \Z} \hat{\chi}^0_{2j+1+2(k+2)l}
  - \sum_{l \in \Z} \hat{\chi}^0_{-2j-1+2(k+2)l} \, .
\eeq
Hence, we obtain
$\hat{\chi}_j \sim \hat{\chi}^0_{2j+1} - \hat{\chi}^0_{-2j-1}$
in the limit of 
$k \rightarrow \infty$, which coincides with the annulus amplitude
\eqref{annulus_W}. 
Since this is the property that characterizes the Ishibashi state
(see \eqref{annulus_Ishibashi}), 
we can identify $\dket{2j+1}^W$ with the $k \rightarrow \infty$
limit of the Ishibashi state $\Ishibashi{j}$.
After this identification, the boundary state 
\eqref{classical_brane} turns out to be the $k \rightarrow \infty$
limit of the Cardy's state, for which the wave function is
the classical one \eqref{classical_wave_function}.
The boundary state \eqref{classical_brane}, 
which realizes the matching condition \eqref{match}, 
correctly reproduce Cardy's state in the limit
$k \rightarrow \infty$. 
 
In order to treat the case of finite $k$, we need to take
into account the quantum effects in the construction of
the boundary states.
We can do this by imposing the invariance of the boundary states
under the Weyl group $\hat{W}$ of $\widehat{su}(2)$ instead
of $W$. 
The Weyl group $\hat{W}$ is a semi-direct product of
the Weyl group $W$ of $su(2)$ and the translation of
the weight lattice.
Hence, the boundary state invariant under $\hat{W}$ 
takes the following form
\beq
  \dket{n}^{\hat{W}} =
    \sum_{l \in \Z} \dket{n  + 2(k+2)l}
  + \sum_{l \in \Z} \dket{-n + 2(k+2)l} \, .
\eeq
Here, we modify the assignment of the picture to be compatible
with the action of the Weyl group
\beq
  \dket{n  + 2(k+2)l} = 
    \begin{cases}
	\dket{n  + 2(k+2)l}_{\varphi} \otimes \dket{0}_{\beta\gamma} & 
	\text{if $0 < n < k+2$} \, , \\
	\dket{n  + 2(k+2)l}_{\varphi} \otimes \dket{-1}_{\beta\gamma} & 
	\text{if $-k-2 < n < 0$} \, . 
	\end{cases}
\eeq
Then the annulus amplitude for $\dket{n}^{\hat{W}}$ gives
the $\widehat{su}(2)$ character $\hat{\chi}_{\frac{n-1}{2}}$:
\beq
  \begin{split}
  {}^{\hat{W}}\dbra{n} 
    q^{\frac{1}{2}(L_0 + \Lt_0 - \frac{c}{24})} 
    z^{\frac{1}{2}(J^3_0 + \Jt^3_0)}
  \dket{n}^{\hat{W}} &= 
  \sum_{l \in \Z} \hat{\chi}^0_{n + 2(k+2)}(\tau, \nu) 
  + \sum_{l \in \Z} \hat{\chi}^{-1}_{-n + 2(k+2)}(\tau, \nu) \\
  &= 
  \sum_{l \in \Z} \hat{\chi}^0_{n + 2(k+2)}(\tau, \nu) 
  - \sum_{l \in \Z} \hat{\chi}^{0}_{-n + 2(k+2)}(\tau, \nu) \\
  &= \hat{\chi}_{\frac{n-1}{2}}(\tau, \nu) \, .
  \end{split}
\eeq
Hence, we can identify the $\hat{W}$-invariant boundary state
$\dket{2j+1}^{\hat{W}}$ with the Ishibashi state $\Ishibashi{j}$.
Since $\dket{n + 2(k+2)}^{\hat{W}} = \dket{n}^{\hat{W}}$,
the normalization factor also should have the periodicity of $2(k+2)$.
The most simple possibility is to take $1/\sqrt{[n]_q}$,
$q = e^{i\frac{2\pi}{k+2}}$, 
instead of $1/\sqrt{n}$.
If we adopt this factor, the quantum boundary state for the $D$-brane
on the conjugacy class takes the following form
\beq
  \dket{a}_{\text{brane}} = \sum_{0 < n < k+2}
  \sin(n a) \frac{1}{\sqrt{[n]_q}} \dket{n}^{\hat{W}} \, .
  \label{quantum_brane}
\eeq
Identifying $\dket{n}^{\hat{W}}$ with the Ishibashi state,
the above state exactly coincides with 
Cardy's state \eqref{Cardy}.

\section{Discussion}
In this paper, we have constructed the boundary state for the spherical
2-brane wrapped around the conjugacy class of $SU(2)$
directly in terms of the group variable.
The resulting state coincides with Cardy's state.
Since we have started from the geometrical setting that 
the worldvolume is a conjugacy class,
this result clarifies the geometrical meaning of Cardy's state.
Our construction is parallel to the one for the $D$-brane in flat 
background.
Namely, we have decomposed the group variable into the left-
and the right-moving coordinates on the worldsheet,
and rewrite the boundary condition for a string ending on
the conjugacy class of $SU(2)$ as the matching
condition of the chiral coordinates.
This formulation of the problem
is preferable to the ordinary boundary condition
since it treats both of the chiral zero modes equally.
In order to write the boundary state subject to
the matching condition, we have used the free field realization 
of the WZW model, 
in which each chiral sector takes values in
the Borel subgroup ($B_+$, $B_-$) of $SL(2,\C)$.
The chiral field $g_L(z)$ ($g_R(\zbar)$) is an upper (lower)
triangular matrix.
We cannot simply impose the
matching condition to the left and right chiral sectors.
Instead, we have used the involution $g\rightarrow(g^\dagger)^{-1}$,
which is trivial in $SU(2)$ but maps $B_+$ to $B_-$. 
With this involution, 
the matching condition for the group variables gives a simple
boundary condition for the free fields.

As we have argued in this paper,
the wave function $\widehat{\psi}^k_\alpha$ for Cardy's state 
takes the following form
\beq
  \widehat{\psi}^k_\alpha(\theta) =
  \frac{S_{\alpha 0}}{\sqrt{S_{00}}} \sum_{j=0, 1/2, 1, \cdots, k/2}
  \sqrt{\frac{2j+1}{[2j+1]_q}} 
  \chi_j(\pi \tfrac{2\alpha + 1}{k+2}) \overline{\chi}_j(\theta) \, .
\eeq
In the limit $k\rightarrow\infty$, this wave function reduces
to the $\delta$-function concentrated on the conjugacy class:
\beq
  \psi^k_\alpha(\theta) =
    \sum_{j = 0, 1/2, 1, \cdots} \!\!
    \chi_j(\pi \tfrac{2\alpha}{k}) \, \overline{\chi_j(\theta)} \, .
\eeq
For finite $k$, however, $\widehat{\psi}^k_\alpha$ is not a
$\delta$-function in the usual sense.
Since the sum is restricted to the integrable representations,
the peak of $\widehat{\psi}^k_\alpha$ is broadened and the worldvolume
of the corresponding $D$-brane gets smeared.
This is because the worldsheet theory is a rational CFT
in which the number of primary fields is finite.
Correspondingly, the Hilbert space of the wave function is
finite dimensional, which suggests that the target space geometry
differs from the ordinary one. 
In ref.\cite{FG,FGR}, the geometry behind a rational CFT is
discussed in the framework of the non-commutative geometry.
Since the worldvolume of the $D$-brane is a submanifold in 
the target space, we expect that the $D$-brane in a rational CFT
also exhibits the feature of the non-commutative geometry.
Actually, it is pointed out \cite{ARS,FFFS} that,
by calculating the algebra of functions on the worldvolume,
the spherical brane in the $SU(2)$ WZW model is 
a quantized sphere,
which reduces to the fuzzy sphere \cite{H,M} 
in the $k \rightarrow \infty$ limit.
Therefore, it is natural to expect that
the above wave function $\widehat{\psi}^k_\alpha$ can be
regarded as a generalization of the ordinary $\delta$-function 
to the case of the non-commutative geometry. 
As we have discussed in the last section,
the factor $[2j+1]_q$ in $\widehat{\psi}^k_\alpha$ is originated from
the change of the normalization of the Ishibashi state 
$\Ishibashi{j}$.
This change of the normalization suggests that the inner product
of the wave function is $q$-deformed, which may also be interpreted
in terms of the non-commutative geometry.
One way to understand the relation between the brane
and the non-commutative geometry is to compare the field
theory on the quantized sphere \cite{GKP,WW}
with the effective field theory on the worldvolume \cite{ARS2}.

Our boundary state is equipped with the structure of the free field
resolution of the irreducible representation by imposing
the invariance under the Weyl group of the current algebra.
In particular, the Weyl reflection exchanges a highest-weight
representation with a lowest-weight one, and we need both of them 
to construct the boundary state invariant under the Weyl reflection.
In the calculation of the character,
the lowest-weight representation behaves like the highest-weight
representation with negative norm, and we reproduce the same structure
as the character formula.
Since our construction is geometrical, it may enable us
to clarify the geometrical interpretation of the BRST cohomology
in the free field realization of the current algebra.

\vskip \baselineskip
\noindent{\Large \bf{Acknowledgement}}
\medskip

The authors would like to thank M.~Kato and U.~Carow-Watamura 
for helpful discussions. 
H.~I. would like to thank also
to H.~Awata, Y.~Satoh, Y.~Sugawara, K.~Sugiyama and S.-K.~Yang
for discussions.
This work is supported by the Grant-in-Aid of Monbusho
(the Japanese Ministry of Education, Science, Sports and Culture) 
\#09640331.


\end{document}